\documentclass[aip,apl,a4paper,reprint,amsmath,amssymb,floatfix]{revtex4-1}
 \usepackage{amsmath}
 \usepackage{bm}
 \usepackage{graphicx}
 \usepackage{SIunits}
 \usepackage{ulem}
 \usepackage{url}
 \usepackage[latin1]{inputenc}

\begin{document}

 \title{High-Q optomechanical GaAs nanomembranes}
 \date{\today}
 \author{J.~Liu}
 \email{jinl@fotonik.dtu.dk}
 \affiliation{DTU Fotonik, Department of Photonics Engineering, Technical University of Denmark, {\O}rsteds Plads 343, DK-2800 Kgs.~Lyngby, Denmark}
 \affiliation{Niels Bohr Institute, University of Copenhagen, Blegdamsvej 17, DK-2100 Copenhagen, Denmark}
 \author{K.~Usami}
 \author{A.~Naesby}
 \author{T.~Bagci}
 \author{E.~S.~Polzik}
 \affiliation{QUANTOP - Danish National Research Foundation Center for Quantum Optics, Niels Bohr Institute, University of Copenhagen, Blegdamsvej 17, DK-2100 Copenhagen, Denmark}
 \author{P.~Lodahl}
 \affiliation{Niels Bohr Institute, University of Copenhagen, Blegdamsvej 17, DK-2100 Copenhagen, Denmark}
 \author{S.~Stobbe}
 \affiliation{Niels Bohr Institute, University of Copenhagen, Blegdamsvej 17, DK-2100 Copenhagen, Denmark}


 \begin{abstract}
      We present a simple fabrication method for the realization of suspended GaAs nanomembranes for cavity quantum optomechanics experiments. GaAs nanomembranes with an area of \unit{1.36}{\milli\meter} by \unit{1.91}{\milli\meter} and a thickness of \unit{160}{\nano\meter} are obtained by using a two-step selective wet-etching technique. The frequency noise spectrum  reveals several mechanical modes in the kilohertz regime with mechanical $Q$-factors up to $2.3\times10^{6}$ at room temperature. The measured mechanical mode profiles agree well with a taut rectangular drumhead model. Our results show that GaAs nanomembranes provide a promising path towards quantum optical control of massive nanomechanical systems.
 \end{abstract}

 \maketitle

    Quantum optomechanics has become an emerging field with the goal of engineering and detecting quantum states of massive mechanical systems such as the quantum ground state\cite{Teufel2011,Wilson-Rae2007,Marquardt2007} or entangled quantum states\cite{Vitali2007,Hammerer2009}. In this rapidly expanding field, various materials and geometries are being pursued, in order to control the coupling between phonons and photons\cite{Kippenberg2008,Marquardt2009,Favero2009}.

    Incorporating direct band gap semiconductor materials in optomechanics exhibits new prospects due to a number of advantageous properties. In particular, GaAs enables the integration of optoelectronic functionality with nanomechanical elements\cite{Ukita1993}. The noncentrosymmetric nature of the zinc-blende crystal structure of GaAs results in an appreciable piezoelectric coefficient, enabling efficient actuation or transduction\cite{Masmanidis2007}. Furthermore, there is a proposal of cooling the lattice temperature of semiconductors down to 10 K, referred to as optical refrigeration, although experimental demonstrations have not yet been realized\cite{Sheik-Hahae2007}. Quantum dots embedded in GaAs also enable strong coupling between a photon and a confined exciton.\cite{Yoshie2004}. Much effort has been invested in fabricating GaAs-based micro-resonators\cite{Cole2008,Cole2010a} and improving their mechanical properties by strain engineering \cite{Yamaguchi2008}. Coupling the intrinsic physical properties of direct band gap semiconductors to mechanical modes would enable a multitude of effects within cavity quantum optomechanics engineering\cite{Midolo2011}. Very recent results on both GaAs disk resonators and InP photonic crystal cavities show the feasibility of realizing optomechanical systems in direct band gap semiconductors\cite{Ding2010,Gavartin2011} but it is widely observed that GaAs microresonators suffer from low $Q$-factors, which limit the optomechanical cooling performance. Here, we present a simple fabrication method for GaAs nanomembranes and demostrate that they exhibit high mechanical $Q$-factors.

    \begin{figure}
    \centering
    \includegraphics[width=\columnwidth]{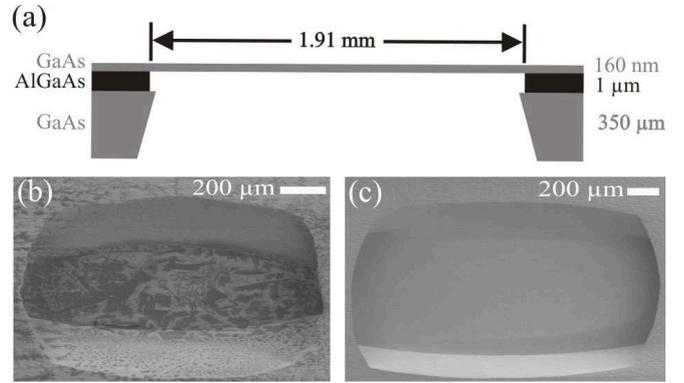}
    \caption{(a) A cross section sketch of the nanomembrane. (b) SEM of the backside of the nanomembrane before and (c) after the cleaning process.}
    \label{fig:SEM}
    \end{figure}

    The GaAs membrane is fabricated from a GaAs/AlGaAs heterostructure wafer comprised of a (100)-oriented GaAs substrate (thickness:~\unit{350}{\micro\meter}), $\mathrm{Al_{\text0.85}Ga_{\text0.15}As}$ etch stop layer (thickness:~\unit{1}{\micro\meter}), and a GaAs capping layer (thickness:~\unit{160}{\nano\meter}), which is shown in Fig.~1(a). The first step is to remove the substrate with selective citric acid wet-etching by using AlGaAs as an etch stop layer\cite{kim1998}. Then a hydrofluoric acid (HF) selective wet-etching follows in order to remove the AlGaAs sacrificial layer\cite{Yablonovitch1989}. The nanomembranes made by our method are optically accessible from both sides, which enables implementing cavity optomechanical cooling schemes.

    The detailed fabrication process is as follows: photoresist with a thickness of \unit{3}{\micro\meter} is coated on both sides of the wafer and the patterns of the holes with different diameters ranging from \unit{500}{\micro\meter} to \unit{1}{\milli\meter} are defined by photolithography. After 30 minutes hard baking at a temperature of $140\ ^\circ$C on a hotplate, the wafer is immersed into citric acid/$\mathrm{H_{2}O_{2}}$ solution (4 Citric acid (50\% by wt.)/1 $\mathrm{H_{2}O_{2}}$ (30\%))\cite{kim1998} with magnetic stirring for 20 hours to etch through the GaAs substrate. Thanks to the excellent etch rate selectivity for GaAs versus AlGaAs in the citric acid/$\mathrm{H_{2}O_{2}}$ solution, the \unit{160}{\nano\meter} GaAs layer is intact due to the protection of the AlGaAs layer and the photoresist during the citric acid wet-etching. Once the substrate is etched away, another selective wet-etching follows to remove the AlGaAs sacrificial layer and finally a GaAs nanomembrane is formed after removing the photoresist.

    \begin{figure}
    \centering
    \includegraphics[width=\columnwidth]{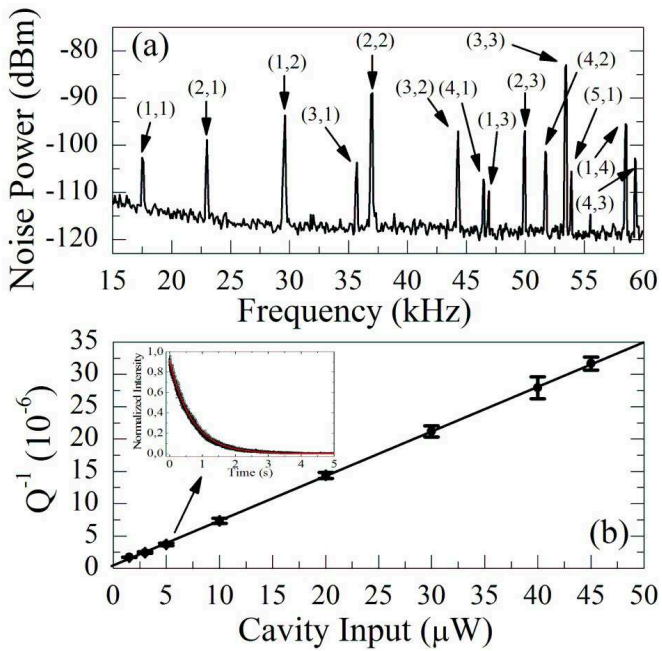}
    \caption{(a) Measured frequency noise spectrum in the 10 kHz - 60 kHz range presenting a series of peaks corresponding to the different mechanical modes labeled by the mode numbers in the parentheses. (b) Inverse $Q$-factor of the (4,3)-mode for different cavity input powers, the points are measurement data and the line is produced by a least-square linear fit. Inset: ringdown measurement with a cavity input power of \unit{5}{\micro\watt}.}
    \label{fig:$Q$factor}
    \end{figure}

    We have observed that often a thin layer of photoresist and submicron-sized particles are left after the completion of photoresist removal, which is shown by the scanning electron micrograph (SEM) in Fig.~1(b). These residues can be significant sources of light scattering and deteriorations of mechanical $Q$-factors. The remnant resist can only be removed by oxygen plasma while the submicron-sized particles speculated to be hydroxide of aluminium\cite{Khankhoje2010a} can be removed in potassium hydroxide (KOH) solutions. Fig.~1(c). shows SEM of a nanomembrane after the oxygen plasma and KOH cleaning processes. We found that the shapes of the fabricated nanomembranes deviated significantly from the circular photolithography masks. This comes from the different etch rates for different crystallographic planes of GaAs in citric acid. A close inspection of the membranes reveals that they are not completely flat, but rather they are bowing with an amplitude at the center of the membranes on the order of a few microns. The size of the largest nanomembrane made by our method is roughly \unit{1.36}{\milli\meter} by \unit{1.91}{\milli\meter}.
    \begin{table}[ht]
    \caption{Mechanical characteristics for the mechanical modes}
    \centering
    \begin{tabular}{c c c c}
    \hline\hline
    Mode & Frequency (kHz) & $Q$-factor ($10^{6}$) & $Q\!\times\! \nu$  ($10^{11}$) \\[0.5ex]
    \hline
    (2,1) & 23.4 & 0.50 & 0.12 \\
    (3,2) & 45.5 & 0.56 & 0.25 \\
    (4,1) & 47.5 & 0.53 & 0.25 \\
    (4,3) & 59.5 & 2.3  & 1.4  \\[0.5ex]
    \hline\hline
    \end{tabular}
    \label{table:mechanical characteristics}
    \end{table}

    We have characterized the mechanical properties of the nanomembranes via frequency noise spectrum by using cavity transmission measurements. A dielectric concave mirror and a 160-nm-thick GaAs membrane form a cavity with the measured finesse of about 10. By feeding the spectrum analyzer with the RF signal from the photodiode, we have observed 14 resonance frequencies ranging from 17.5 kHz (the (1,1)-mode) to 59.5 kHz ((4,3)-mode), see Fig.~2(a). The $Q$-factor is defined as $Q$ = $\omega/\Gamma$ where $\nu = \omega/2\pi$ is the resonance frequency and $\Gamma$ is the mechanical damping rate. $\Gamma$ is measured through a ringdown measurement in which the ringdown of the mechanical modes is observed by inducing the oscillations with an intensity-modulated cavity field. A typical ringdown measurement for the (4,3)-mode is shown in the inset of Fig.~2(b). Due to the optomechanical effect\cite{Kippenberg2008,Marquardt2009,Favero2009} in the ringdown measurements where the $Q$-factors and resonance frequencies vary with input powers, the intrinsic mechanical $Q$-factor of the (4,3)-mode is extracted from the power-dependent ringdown measurement in Fig.~2(b). By extrapolating the $Q$-factor data to zero cavity input power, a $Q$-factor of $2.3\times10^{6}$ is found at 59.5 kHz. Since the damping rate of a mechanical mode (i.e., inverse of Q) generally scales linearly with $\nu$, $Q\times\nu$ is an important parameter for the mechanical characterizations.  Table 1 lists the resonance frequencies, the corresponding $Q$-factors and the product of $Q$ and $\nu$ for the modes we have carefully characterized. The highest $Q\times\nu$ value in our case is $1.4\times10^{11}$, which is comparable to the state of art for GaAs resonators at cryogenic temperatures, cf. Ref.~\onlinecite{Yamaguchi2008}. We stress that much higher $Q$-factors for our GaAs nanomembranes are expected at cryogenic temperatures.

    \begin{figure}
    \centering
    \includegraphics[width=\columnwidth]{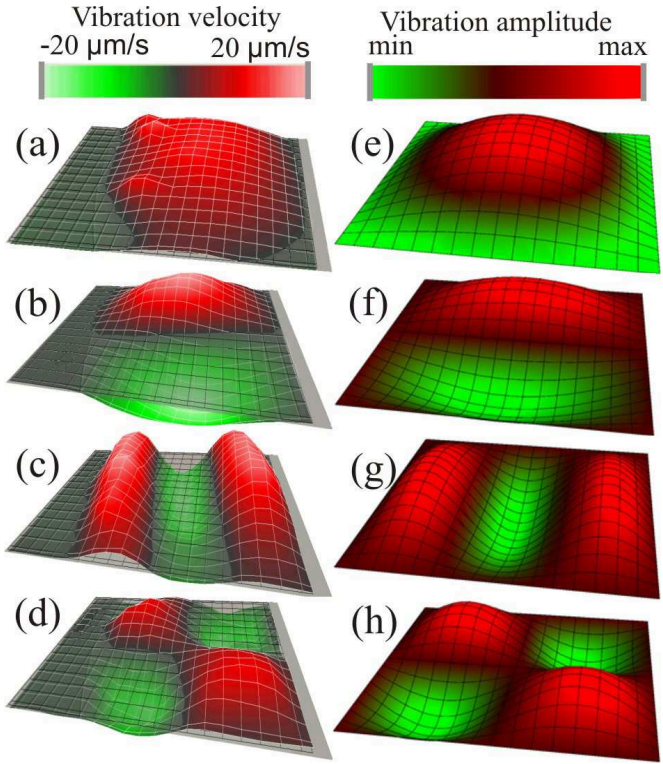}
    \caption{Analysis of the experimental and computed mechanical mode profiles. (a)-(d): Measured deformation profiles of the (1,1), (1,2), (3,1), and (2,2)-mode respectively with a micro system analyzer. (e)-(h): calculated mechanical mode profiles of the (1,1), (1,2), (3,1), and (2,2)-modes respectively according to a taut rectangular drumhead model.}
    \label{fig:Modeprofiles}
    \end{figure}

    The vibration amplitudes, which are proportional to the vibration speeds for a given resonance frequency, are mapped by a laser-Doppler vibrometer [MSA-500 Micro System Analyzer (Polytec Ltd)] with a white light interferometry method. Due to space limitations in the experimental method, these measurements were performed on a smaller sample (roughly \unit{0.7}{\milli\meter} by \unit{1.4}{\milli\meter} ) than the one presented in Fig.~1(c).  The vibration profiles of the (1,1), (1,2), (3,1), and (2,2)-modes are shown on the left-hand side of Fig.~3 while the corresponding numerical simulations are on the right-hand side. The simulated vibration modes are based on a taut rectangular drumhead model. We find a good agreement between measured mechanical mode profiles and simulations. This model is also used to estimate the fundamental resonance frequency. The internal stress of the Al$_{x}$Ga$_{1-x}$As/GaAs interface due to the lattice mismatch during growth can be approximated as\cite{Hjort1994157} $133\times$$x$ and in our case $x$ = 0.85. Thus a tensile stress of 110 MPa is predicted. By including this stress, we get a fundamental resonance frequency (the (1,1)-mode) of 63 kHz which is higher than the measured resonance frequency. More investigations are needed to obtain a better understanding of the relation between the stress and high $Q$-factors in this system.

    In conclusion, we have realized high-$Q$ GaAs nanomembranes with unprecedented sizes for cavity quantum optomechanics experiments. The resonance frequencies and mode profiles are both theoretically and experimentally investigated. Experimental data show good agreement with a simplified model based on a taut rectangular drumhead. The ringdown measurements reveal that the highest $Q$-factors of the mechanical modes can be up to $2.3\times10^{6}$ at room temperature, which underlines the potential of these nanomembranes for cavity cooling experiments. We will report on such studies in a future publication. Further steps can be envisioned by incorporating semiconductor quantum dots in nanomembranes to cool the nanomechanical modes to its ground state\cite{Wilson-Rae2004} and patterning the nanomembranes for controlling of the coupling between phonons and photons\cite{Alegre2010}.

    We gratefully acknowledge S.~Schmid and A.~Boisen for the assistance of vibration profile measurements and financial support from the Villum Kann Rasmussen Foundation, The Danish Council for Independent Research (Natural Sciences and Technology and Production Sciences), the European Research Council (ERC consolidator grant), and the European project Q-ESSENCE.

    \end{document}